\documentclass{article}

\usepackage{graphicx}
\usepackage{psfig}
\usepackage{epsfig}
\usepackage[round]{natbib}

\title{Do medium range ensemble forecasts give useful predictions of temporal correlations?}

\author{Stephen Jewson\footnote{\emph{Correspondence address}: RMS, 10 Eastcheap,
London, EC3M 1AJ, UK. Email: \texttt{x@stephenjewson.com}}\\
RMS, London, United Kingdom}
\begin{document}

\newcommand{\bx}[1]{\fbox{\begin{minipage}{15.8cm}#1\end{minipage}}}

\maketitle
\begin{abstract}
Medium range ensemble forecasts are typically used to derive
predictions of the conditional marginal distributions of future events on
individual days. We assess whether they can also be used to
predict the conditional correlations between different days.
\end{abstract}

\section{Introduction}

We consider the question of how to make probabilistic forecasts of
future temperatures over the 1-10 day timescale. A complete
specification of the distribution of future temperatures over this
time period would consist of information about the marginal
distributions of temperature on each day and about the
dependencies between the temperatures on different days. 
Investigating these marginal distributions and the dependency structure
in full generality is rather difficult, and so we will
make the approximation that temperature is normally distributed.
This simplifies the problem greatly since
in this case the distribution of future temperatures is described
completely by 10 means, 10 variances, and a 10 by 10 correlation
matrix.

Ensemble forecasts can be used to derive predictions of the mean,
and such predictions are better than predictions
derived from single integrations of forecast models at the same resolution.
Ensemble forecasts can also be used to derive predictions of the
variance. There are a number of ways this can be done. We have
analysed some of these in detail in previous articles, 
including using linear regression on the ensemble mean, and spread regression on the
ensemble mean and spread 
(see~\citet{jewsonbz03a}, \citet{jewson03i}, \citet{jewson03h} and \citet{jewson03g}).
We have only been able to show that the
spread of the ensemble improves the skill of forecasts when the calibrated forecasts are
evaluated in-sample (i.e. the calibration and evaluation are performed on the same data that
is used to calculate the calibration parameters).
For out of sample forecasts it seems that it is very
hard to prove that one can beat linear regression on the ensemble mean as a calibration method. 
We argue that this is because the predictable part of the variability in the variance is small and
because only short records of past forecasts are available for training 
calibration models.

We now address the question of whether ensemble forecasts can be
used to predict the correlations between temperatures on different days
of the forecast, or whether such predictions should be made using
past-forecast error statistics. The potential advantage of
deriving the prediction of correlations from the ensemble is that
it will then be flow-dependent. It is highly plausible that as the
atmospheric state changes the correlations between forecast errors
on different days of the forecast should change, and ensembles have the
potential to predict that effect whereas past forecast error statistics do not. 
On the other hand numerical model forecasts are very prone to biases and our
experience with trying to predict the variance of forecast errors
from the ensemble has taught us that nothing from an ensemble
forecast should be taken for granted: everything needs careful
analysis and calibration to extract the useful information.

The author became interested in the question of whether ensembles can
predict temporal correlations because it arises in a simple
weather derivative pricing situation, as described in~\citet{jewsonc03b}.
Consider a weather option
based on December mean temperature. To calculate the fair value of
such an option one has to estimate the distribution of the
settlement index i.e. the distribution of December mean
temperatures. When it is estimated many months prior to the start
of the contract, this distribution can be derived entirely from
historical data, while when it is estimated immediately prior to or during
the contract it should be derived from a combination of historical
data and forecasts.
For example, imagine that we are estimating this
distribution on the 1st of December. The expectation of the
settlement index is then given by the sum of two values. The first of these values
is the contribution from Dec 1st to Dec 10th (which can be estimated from
a forecast) and the second is the contribution from Dec 11th to Dec 31st
(which can be estimated from historical data).
Thus estimating the mean
of the settlement index is rather straightforward.
Estimating the variance of the
index, however, is more complicated, and involves making estimates of
the variances of temperatures on each day of
the month (31 values) and the correlations between temperatures on different
days of the month (a 31x31 element matrix). 
One part of estimating this 31x31 matrix is to estimate the 
correlations between the days of the forecast (a 10x10 element sub-matrix) and this is what
motivates the question of whether those correlations can be
derived from the ensemble. A more general discussion of how the
rest of the correlations and variances in this problem can be
calculated is given by~\citet{jewsonc03b}.

\section{Data}

We will base our analyses on one year of ensemble forecast data for the weather
station at London's Heathrow airport, WMO number 03772. The forecasts are predictions
of the daily average temperature, and the target days of the forecasts
run from 1st January 2002 to 31st December 2002. The forecast was produced
from the ECMWF model~\citep{molteniet96} and downscaled to the airport location using a simple
interpolation routine prior to our analysis. There are 51 members in the ensemble.
We will compare these forecasts to the quality controlled climate
values of daily average temperature for the same location as reported by the UKMO.

There is no guarantee that the forecast system was held constant throughout this period,
and as a result there is no guarantee that the forecasts are in any sense stationary,
quite apart from issues of seasonality. This is clearly far from ideal with respect to 
our attempts to build statistical interpretation models on past forecast data but is,
however, unavoidable: this is the data we have to work with.

Throughout this paper all equations and all values are in terms of double anomalies
(have had both the seasonal mean and
the seasonal standard deviation removed). 
Removing the seasonal standard deviation
removes most of the seasonality in the forecast error statistics, and partly justifies the use of
non-seasonal parameters in the statistical models for temperature that we propose.

\section{Models}

There are potentially many ways that one could address the
question of whether or not the ensemble can be used to predict
temporal correlations. Since this question has not, apparently,
been addressed before, we will take a simple and pragmatic
approach, which works as follows.

We will model the mean and variance of the forecast using the
spread regression model of~\citet{jewsonbz03a} and will perform in-sample
calibration of all the forecasts using this model. 
We are happy to perform this mean-variance calibration entirely in-sample
because we are addressing
the question of whether the \emph{correlations} contain useful information, rather
than whether the mean and variance of the ensemble do.
We note that this calibration does not affect the correlations between days. 

We will model the correlation matrix between the days of the
forecast as a weighted sum of the two matrices $C^{\mbox{past forecast error statistics}}$
and $C^{\mbox{ensemble forecast}}$ as follows:

\begin{equation}\label{matrix}
  C_i=\lambda C^{\mbox{past forecast error statistics}}+ (1-\lambda)C_i^{\mbox{ensemble forecast}}
\end{equation}

where $C_i$ is the modelled correlation matrix on day $i$, 
$C^{\mbox{past forecast error statistics}}$ is a
stationary matrix derived from past forecast error statistics, 
and
$C_i^{\mbox{ensemble forecast}}$
is a time-varying matrix derived from the ensemble
forecast. We will vary the weighting $\lambda$ of these matrices from zero to one and derive
the combination that gives the optimum probabilistic forecast,
defined as that forecast which maximises the likelihood. 

Our first test is an in-sample test that fits the past forecast error based correlation
matrix on the whole year of data, and then tests it on the
same year of data. One can argue that this test is not
very useful, since
we are fitting a very large number of parameters (the 56
independent elements of the correlation matrix) on a relatively
small amount of data. There is a very large danger of over-fitting.

Our second test avoids this problem by fitting the
past forecast error based correlation
matrix on the first six months of data and testing it on the second
six months of data, and vice versa.

The likelihood score that we will attempt to maximise by combining
the two correlation matrices is given by the multivariate normal
distribution over all forecast days and all leads. We will assume
that forecasts are independent from day to day but not from lead to lead.
The likelihood then becomes:

\begin{equation}\label{likelihood}
    L=\prod_{i=1}^{i=N}\frac{1}{(2\pi)^\frac{n}{2}D_i^\frac{1}{2}}\mbox{exp} \left( -\frac{1}{2}(e_i^T C_i^{-1} e_i) \right)
\end{equation}

where $N$ is the number of days of forecasts, 
$n$ is the number of forecast leads, $e_i$ is a vector of
forecast errors on day $i$ (of length $n$), $C_i$ is the estimated forecast
error covariance matrix on day $i$ (of dimension $n$ by $n$) as given by equation~\ref{matrix},
 and $D_i$ is the determinant of this matrix.

\section{Results}

\begin{figure}[!htb]
  \begin{center}
    \includegraphics{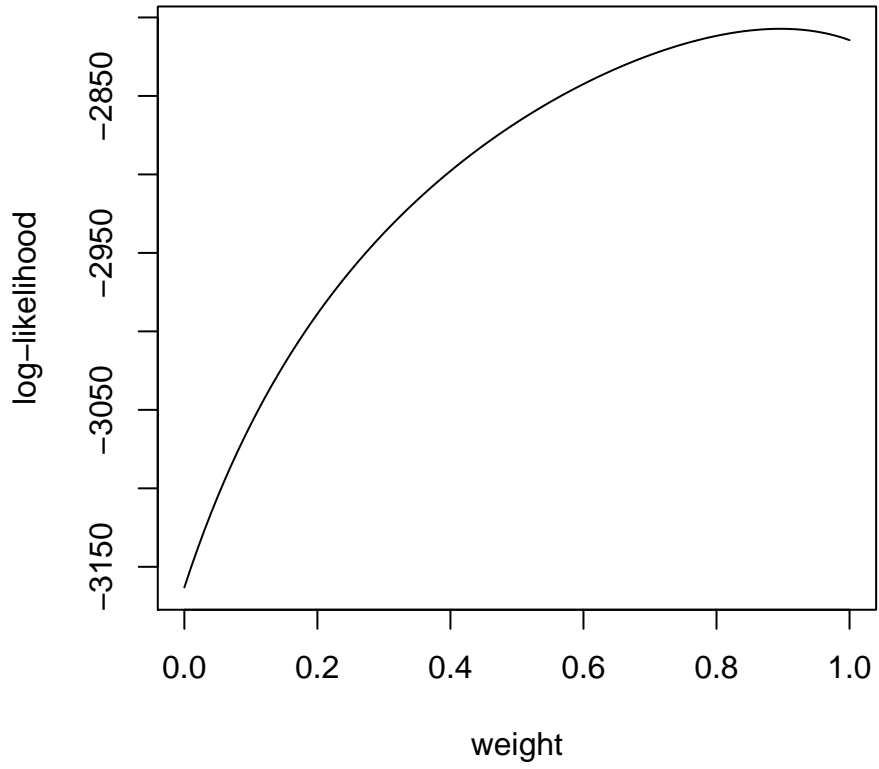}
  \end{center}
  \caption{
    The log-likelihood for one year of forecast data versus the weighting used 
    to derive the inter-lead correlation matrix.
          }
  \label{f:f1}
\end{figure}

Our in sample results are shown in figure~\ref{f:f1}. The
horizontal axis shows the weight applied to the past forecast error
based correlation matrix (the $\lambda$ in equation~\ref{matrix}). 
We have plotted the log-likelihood, given by the log of
the likelihood from equation~\ref{likelihood}. As we vary the
correlation matrix, the likelihood changes. We see that the
highest values for the likelihood are given when we weight the two
correlation matrices roughly in the proportions 90\% (for the
past forecast error based matrix) to 10\% (for the ensemble forecast based matrix). In
spite of the caveats we have about the in-sample nature of this
test, the results are somewhat interesting in that they certainly imply
that the ensemble forecast based correlation matrix contains some useful
information. Because it has been performed in sample we would
expect this test to strongly favour the past forecast error based matrix,
and hence to be biased towards higher values of $\lambda$ than
out of sample tests.

\begin{figure}[!htb]
  \begin{center}
    \includegraphics{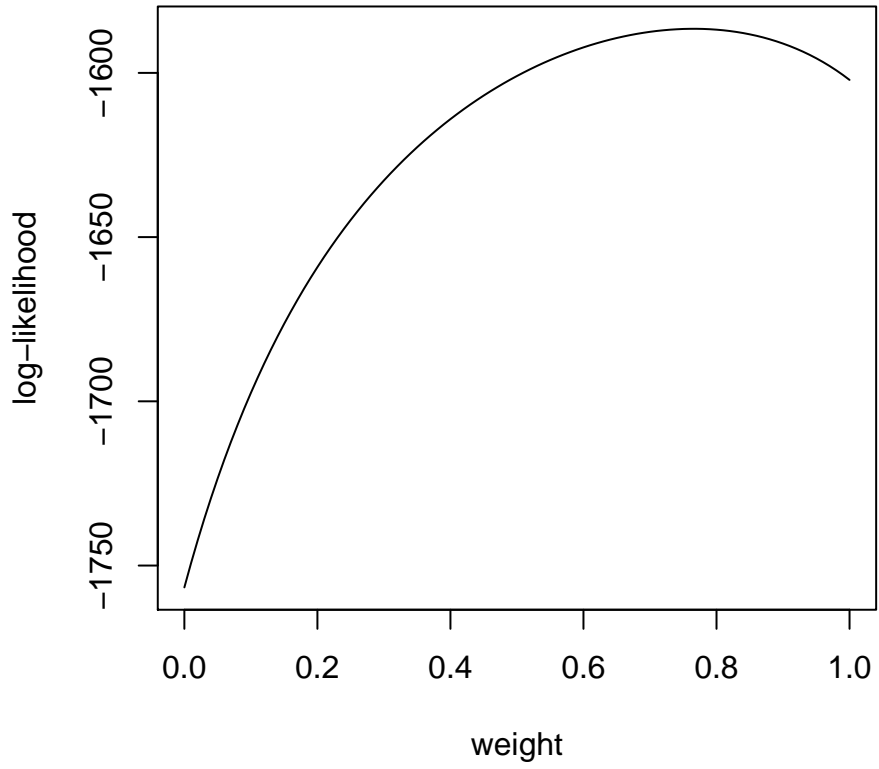}
  \end{center}
  \caption{
    The log-likelihood for 6 months of forecast data versus the weighting used to
    derive the inter-lead correlation matrix. In this case the past forecast error
    based component of the correlation matrix was calculated on a different 6 month
    data period from that used to calculate the log-likelihood.
          }
  \label{f:f2}
\end{figure}

\begin{figure}[!htb]
  \begin{center}
    \includegraphics{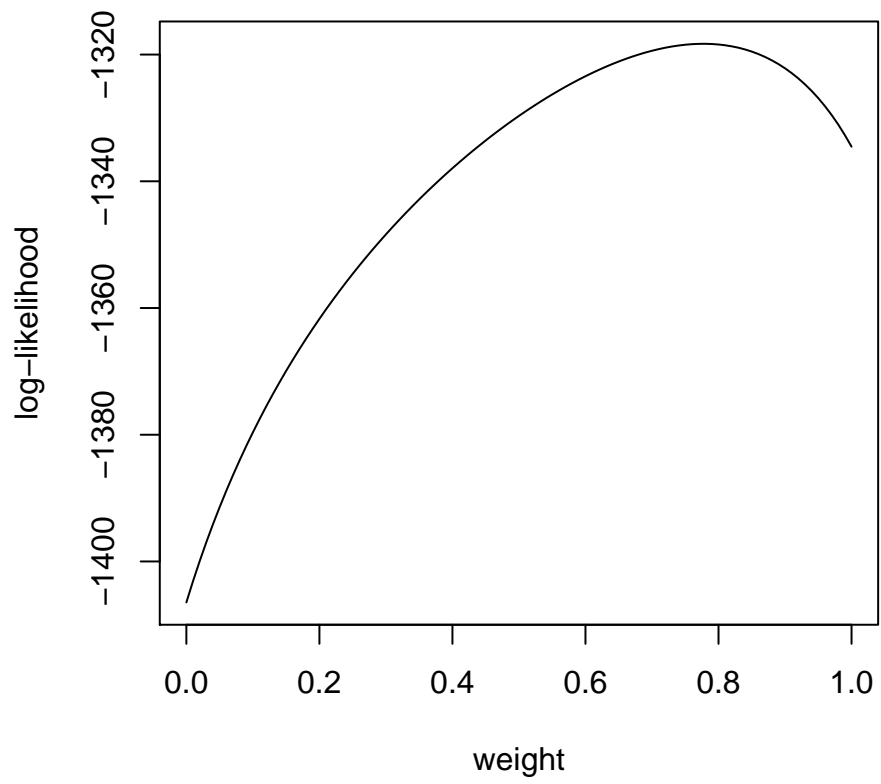}
  \end{center}
  \caption{
    As figure~\ref{f:f2} but with the data periods exchanged.
          }
  \label{f:f3}
\end{figure}

The out of sample results, in which the data being predicted is
different from the data used to calculate the past forecast error correlation
matrix, are shown in figures~\ref{f:f2} and~\ref{f:f3}. 
We see that in both cases the likelihood has a maximum at around
80\% (in fact, the exact numbers are 77\% and 78\%). As we expected,
the optimum combination is at a lower level for the weight than for the in-sample tests.
The results for the two tests are remarkably consistent, giving us
reasonable faith that sampling error is not playing too important a role.

If we were forced to use either the past forecast error based correlation
matrix or the ensemble based correlation matrix, then we see clearly
that the past forecast error based correlation matrix performs better i.e. gives
higher values for the log-likelihood.
However there is a wide range of linear combinations of the two matrices that
performs better still. 

\begin{figure}[!htb]
  \begin{center}
    \includegraphics{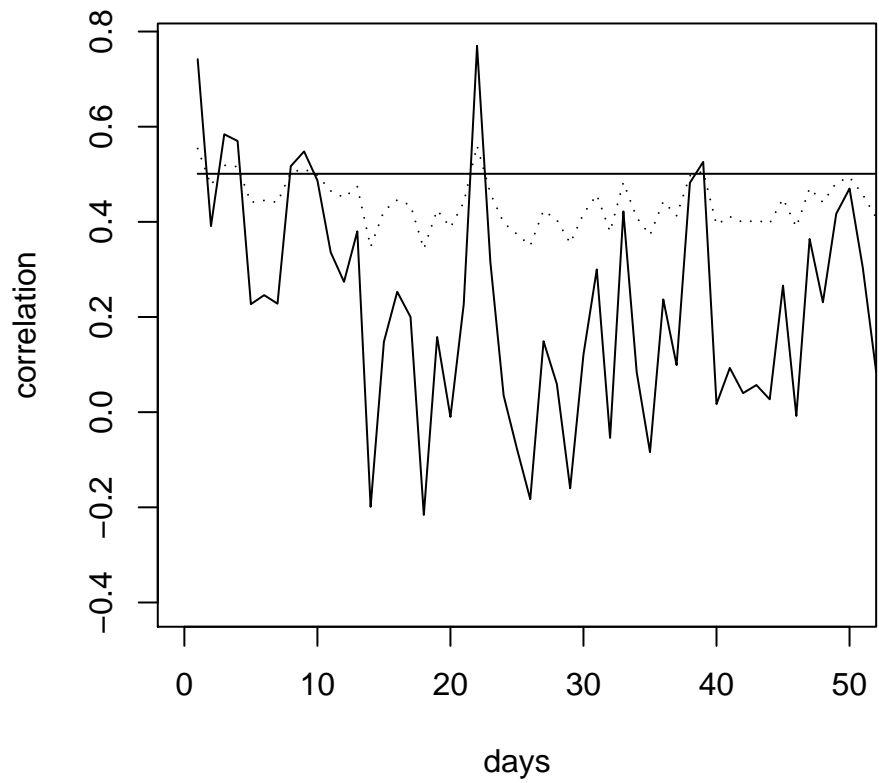}
  \end{center}
  \caption{
    An example of the values for the inter-lead correlations, showing the correlations
    between lead 2 and lead 3. The solid lines show the correlations
    based on an ensemble forecast (varying line) and on past forecast error statistics (constant values), 
    and the dotted line shows the optimum combination of these two.
          }
  \label{f:f4}
\end{figure}

In figure~\ref{f:f4} we show an example of the correlation time
series derived from the optimum values of lambda. This example is based on data
from the first 50 days of the predicted data from the first of the out of 
sample tests, and shows the correlation between leads 2 and 3.
The two solid
lines show the correlations from the past forecast errors and the ensemble
forecast, while the dotted line shows the weighted combination of
these two correlations using the optimum value for lambda. We see that
the ensemble based correlation is on average lower than the correlation based on 
past forecast errors, but that for some values it is higher. The optimum
weighting of the two reduces the variability of the predicted correlation
very dramatically.

\section{Summary}

We have investigated whether the temporal correlations derived
from an ensemble forecast are useful predictors for the
correlations between the distributions of possible temperatures on
each day of the forecast. We find:

\begin{itemize}
    \item when we compare probabilistic forecasts derived from a correlation matrix
    based on past forecast error statistics with forecasts derived
    from a correlation matrix based on the ensemble, those based
    on past forecast error statistics are better
    \item a linear combination of the two
    correlation matrices performs better than either matrix on its
    own
    \item the optimum proportions of the two matrices seem to be
    fairly robustly given by around 77\% for the past forecast error
    correlation matrix and around 23\% for the ensemble derived
    correlation matrix
\end{itemize}

We have deliberately used a very simple methodology. In
particular, we have used just a single weight for the whole
correlation matrix. One could also consider using different weights for
the different elements of the matrix. This might give better results, since one
could certainly imagine that the ensemble would give better
correlations at shorter leads, while the past forecast error based correlations
would perform better at longer leads. If having separate weights
for each member of the matrix proves unwieldy (there would be 56
different weights) one could consider parametrising the structure
of the weights with a smaller number of parameters.

In terms of further work, there are two other questions 
related to the predictability of correlations
that immediately spring to mind. The first is to ask: do
ensemble forecasts contain useful correlation information about
the correlation \emph{between} stations? One could look at both
the instantaneous cross-correlations, and the lag
cross-correlations. The second is: do ensembles contain useful
information about the correlation between different variables i.e.
between temperature and precipitation at the same location and the same lead
time, between temperature and precipitation at a different
location but the same lead time, or even between temperature and
precipitation at a different location and at a different lead
time.

\section{Acknowledgements}

Thanks to Ken Mylne for providing the data and Rie Kondo for her patience.

\bibliography{temporalcorrelation}

\newpage

\newpage

\newpage

\newpage

\end{document}